\documentclass{nature}
\usepackage{graphicx}
\usepackage{amsfonts,amsmath,amscd,amssymb} 
\usepackage{hyperref}
\usepackage{bibunits}
\usepackage{xpatch}
\usepackage[switch, modulo]{lineno}
\modulolinenumbers[1]

\usepackage[dvipsnames]{xcolor}

\newcommand{\Msun}{M$_\odot$}

\newcommand{\gppr}{\stackrel{>}{\scriptstyle \sim}}
\newcommand{\gappr}{\raisebox{-0.4ex}{$\gppr$}}

\title{The origin and evolution of magnetic white dwarfs in close binary stars}

\author{
Matthias R. Schreiber$^{1,2,*}$,
Diogo Belloni$^{3,1,*}$, 
Boris T. G\"ansicke$^{4}$,
Steven G. Parsons$^{5}$,
Monica Zorotovic$^{6}$
}

\begin{document}

\begin{bibunit}

\maketitle

\begin{affiliations}
\item Departamento de F\'isica, Universidad T\'ecnica Federico Santa Mar\'ia, Av. España 1680, Valpara\'iso, Chile 
\item Millennium Nucleus for Planet Formation (NPF), Valpara\'iso, Chile 
\item National Institute for Space Research, Av. dos Astronautas, 1758, 12227-010, S\~ao Jos\'e dos Campos, SP, Brazil
\item Department of Physics, University of Warwick, Coventry CV4 7AL, UK
\item Department of Physics and Astronomy, University of Sheffield, Sheffield S3 7RH, UK
\item Instituto de F\'isica y Astronom\'ia, Universidad de Valpara\'iso, Av. Gran Bretaña 1111, Valpara\'iso, Chile\\
$^*$These authors contributed equally to this work
\end{affiliations}

\begin{abstract}
The origin of magnetic fields in white dwarfs remains a fundamental unresolved problem in stellar
astrophysics. In particular, the very different fractions of strongly (\gappr\,1 MG) magnetic white dwarfs in
evolutionarily linked populations of close white dwarf binary stars cannot be reproduced by any
scenario suggested so far. Strongly magnetic white dwarfs are absent among detached white dwarf
binary stars that are younger than approximately 1 Gyr. In contrast, in semi-detached cataclysmic
variables in which the white dwarf accretes from a low-mass star companion, more than one third
host a strongly magnetic white dwarf\cite{palaetal20-1}. Here we present binary star evolutionary
models that include the spin evolution of accreting white dwarfs and crystallization of their cores, as
well as magnetic field interactions between both stars. We show that a crystallization- and rotation-driven
dynamo similar to those working in planets and low-mass stars\cite{christensenetal09-1} can
generate strong magnetic fields in the white dwarfs in cataclysmic variables which explains their large
fraction among the observed population. When the magnetic field generated in the white dwarfs
connects with that of the secondary stars, synchronization torques and reduced angular momentum
loss cause the binary to detach for a relatively short period of time. The few known strongly magnetic
white dwarfs in detached binaries, including AR Sco\cite{marshetal16-1}, are in this detached phase.
\end{abstract}

\noindent
The vast majority of close binary stars containing at least 
one white dwarf form through common envelope evolution\cite{ivanovaetal13-1}. 
The emerging detached post common envelope binary stars evolve towards shorter 
orbital periods driven by angular momentum loss\cite{schreiber+gaensicke03-1} and 
eventually become semi-detached cataclysmic variable stars (CVs)\cite{kniggeetal11-1}. 
Despite this clear evolutionary link, the fraction of strongly magnetic white dwarfs 
differs drastically between both types of close white dwarf 
binaries. 

We know more than 160 CVs with a strongly
(larger than $1$\,MG) magnetic white dwarf\cite{ferrario15-1}.
In most of these systems the white dwarf accretes 
from its Roche-lobe filling low-mass star companion along the magnetic field lines. 
Due to interactions of the magnetic fields of both stars, the rotation of the 
white dwarf is synchronised with the orbital motion 
of the secondary star, such systems are called {\em{polars}}. 
In a smaller but still important fraction of CVs, the magnetic field of the white dwarf disrupts the inner parts of the accretion disk but the magnetic fields of both stars do not connect. In these so-called {\em{intermediate polars}} the rotation of the white dwarf and the orbital motion are therefore not synchronised\cite{nortonetal04-1}. 
A recent study of a volume-limited sample has shown that over a third of all CVs contain a magnetic white dwarf, with polars and intermediate polars making up 
approximately~$29$ 
and $7$ per cent of the total population\cite{palaetal20-1}.

The situation is very different among the detached systems that are believed to be the progenitors of CVs.  
Only 15
strongly magnetic white dwarfs in close detached binaries are currently known\cite{schwopeetal09-1,parsonsetal20-1}
and they make up less than two per cent of the over one thousand known systems\cite{schreiberetal10-1,rebassa-mansergasetal12-1}. 
All but one of these detached magnetic white dwarf binaries have 
been identified
due to a peculiar emission line\cite{reimersetal99-1}, which turned out to be the third harmonic of a cyclotron fundamental emitted by a low-density plasma. This observational finding 
is convincingly interpreted as resulting from wind accretion onto a strongly 
magnetic white dwarf in a close but detached binary\cite{schmidtetal05-1}. Because angular 
momentum loss 
will eventually drive them into a semi-detached CV configuration, these systems
have been termed {\em{pre-polars}}\cite{schwopeetal09-1}. 
The only exception is AR\,Sco, which was discovered because of the optical and radio pulses of its rapidly ($1.97$\,min) rotating white dwarf\cite{marshetal16-1}.
AR\,Sco and most pre-polars are located in the orbital 
period range between 3 and 5 hours,  
their secondary stars are close to filling their Roche-lobes 
($R/R_{\mathrm{L}}$ typically exceeding $80$ per cent)\cite{parsonsetal20-1}, 
and the magnetic white dwarfs are cold with effective temperatures typically below 10,000\,K 
which implies that they formed at least one Gyr ago.  
Among younger close detached white dwarf binaries, not a single strongly magnetic white dwarf has been found\cite{liebertetal05-1,belloni+schreiber20-1}. 
This puzzling situation, 
a large fraction 
(approximately $35$ per cent) 
of strongly magnetic white dwarfs in CVs
and the much smaller fraction of magnetic white dwarfs 
among the detached systems that are believed to be CV progenitors, was aptly expressed by 
Liebert\cite{liebertetal05-1} who asked
{\em{Where Are the Magnetic White Dwarfs with Detached, Non-degenerate Companions?}}
This fundamental question of stellar evolution has remained without 
an answer for 15 years 
and the solution must link the evolution of white dwarf binaries and
magnetic field generation in white dwarfs. 

Several hypotheses have been put forward to explain 
magnetic field generation in white dwarfs in the last decades. 
In their present form, the fossil field scenario\cite{braithwaite+spruit04-1}, a dynamo operating during common envelope 
evolution\cite{toutetal08-1}, 
or coalescing double white dwarfs\cite{garcia-berroetal12-1} 
face one serious problem when considered as the main formation mechanism for close magnetic white dwarf binaries. In all 
these cases the magnetic fields are generated during the formation process of the white dwarf, i.e. according to the current versions of these theories also detached post common envelope binaries containing hot white dwarfs should host strong magnetic fields which is clearly not the case\cite{belloni+schreiber20-1}.

As the only mechanism that depends on the age of the 
white dwarf, a dynamo similar to those operating 
in low-mass main-sequence stars 
and planets\cite{christensenetal09-1} 
has been suggested to operate if the white dwarf is rotating rapidly 
when the core of the white dwarf is crystallizing\cite{isernetal17-1}. As a white dwarf with a Carbon/Oxygen core cools,
the ions in the core begin to freeze in a lattice structure. 
During crystallization the chemical potential, temperature, and pressure of the liquid and solid phases must remain equal. 
As a consequence, in crystallizing white dwarfs   
the solid phase becomes richer in oxygen and sinks while the carbon excess mixes with the outer liquid envelope which is redistributed by Rayleigh–Taylor instabilities. This configuration is similar to that found in the core of the Earth, where the light element release associated with the inner core growth is a primary driver of the dynamo generating the Earth's magnetic field\cite{lister+buffet95-1}. 

The predicted field strengths generated by this dynamo in white dwarfs depend entirely on the scaling law that is assumed. Applying the same 
scaling law that is 
used for planetary magnetic fields leads to field strengths in white dwarfs of up to 
approximately~$1$\,MG\cite{isernetal17-1}. 
However, taking into account a likely dependence of the dynamo efficiency on the 
magnetic Prandtl number
and that in white dwarfs this number is orders of magnitudes larger than in planets, one could expect this dynamo to be able to generated 
fields much stronger than $1$\,MG (for more details see Methods). 

Assuming that the crystallization 
and rotation driven dynamo can generate 
strong magnetic fields we computed evolutionary tracks starting with non-magnetic 
post common envelope binaries.  
We incorporated several physical mechanisms in the 
stellar evolution code MESA\cite{paxtonetal19-1}.  
The spin-up of the white dwarf due to the accretion of angular momentum\cite{kingetal91-1} was included as well as
crystallization of the white dwarf\cite{bedardetal20-1}.
We furthermore incorporated the reduction in angular momentum loss through magnetic 
braking when the magnetic fields of both stars connect\cite{bellonietal20-1}, a 
condition for synchronisation\cite{hameuryetal87-1}, and
a prescription for angular momentum transfer from the spin of the white dwarf 
to the orbit 
during synchronisation (see Methods for more details). 

Figure\,1 shows an exemplary evolutionary track resulting from our simulations and a schematic illustration of the different stages the system passes through. 
In our model the white dwarf is born without a strong magnetic field 
which is in agreement with the absence 
of strongly magnetic white dwarfs in young close detached post common envelope binaries (stage 1). 
The non-magnetic detached binary star then evolves slowly towards shorter orbital periods
and the white dwarf cools. 
When the secondary fills its Roche-lobe the binary becomes a CV, accretion spins up the white dwarf\cite{kingetal91-1} and the core may be crystallizing (stage 2). If this happens, both conditions for the crystallization and rotation driven 
dynamo are met and a strong magnetic field is generated (stage 3). 
If the magnetic field is strong enough to connect with the field of 
the secondary star the latter provides a synchronising torque on the white dwarf spin. 
The spin angular momentum transferred to the orbital motion due to the synchronising 
torque causes the secondary to detach from its Roche-lobe. This effect is supported by 
the fact that the connection of the two magnetospheres reduces angular momentum loss due to magnetic wind braking because part of the gas escaping from the secondary is accreted by the white dwarf along the connected field lines, instead of leaving the system\cite{bellonietal20-1}.
At the beginning of synchronisation the system becomes a radio pulsing white dwarf binary, such as AR\,Sco, 
in which a rapidly rotating white dwarf spins down 
on a relatively short 
time scale (stage 4) of approximately $1-10$\,Myr.
When the white dwarf spin is synchronised with the orbital motion of the secondary star, the binary star has evolved into the pre-polar stage in which wind accretion generates the observed cyclotron emission (stage 5).
Angular momentum loss through reduced magnetic braking 
and gravitational radiation brings the stars closer together until the secondary 
again fills its Roche-lobe and finally a polar is formed (stage 6). 
The detached phase (stage 3-5) occurs preferentially for orbital periods shorter than  
roughly~$5$\,hr 
as the reduction of magnetic braking and the transfer of spin angular momentum to the orbit have less
impact at longer periods. 
Detached phases occur less frequently at orbital periods shorter than 
approximately~$3$\,hr 
because the conditions for magnetic field generation are typically met at early stages of CV evolution (see Methods for a detailed comparison of observations and model predictions). 

While a different mechanism
such as common envelope merger\cite{toutetal08-1} and/or double white dwarf merger\cite{garcia-berroetal12-1}  
is probably required to explain most single white dwarfs with magnetic field strengths exceeding 1\,MG (see Supplementary Material for details),
the new evolutionary sequence predicted by our model is in excellent agreement with the observations of close binary stars containing 
strongly magnetic white dwarfs.
As the conditions for the dynamo are relatively frequently met in CVs (see Supplementary Material for a discussion of the predicted 
relative numbers), 
our model successfully reproduces the large fraction of CVs containing magnetic white dwarfs. 
The few known detached binaries containing magnetic white dwarfs are 
not progenitors of CVs but instead systems that have been non-magnetic CVs in the past and where mass transfer stopped because the field was generated. This naturally explains why there is not one single hot and young strongly magnetic white dwarf in a detached binary. 
Instead our model predicts that all of them 
are older than 1 Gyr with crystallizing cores as is observed (Fig.\,2) and also explains why all 15 currently known close detached binaries containing magnetic white dwarfs are close to Roche-lobe filling (Supplementary Figure\,1).
Furthermore, the observed clustering of 
pre-polars in the orbital period range of $3-5$\,hr is well reproduced in our simulations as the detached phase typically occurs in this range. 
Finally, our evolutionary sequence predicts short phases of rapidly rotating and strongly
magnetic white dwarfs in detached binaries 
thereby offering an explanation for the existence of the 
radio pulsating system AR\,Sco. We conclude that a dynamo-driven generation of magnetic fields in white dwarfs resolves several long standing problems at once. 


\clearpage


\centerline{\includegraphics[width=150mm]{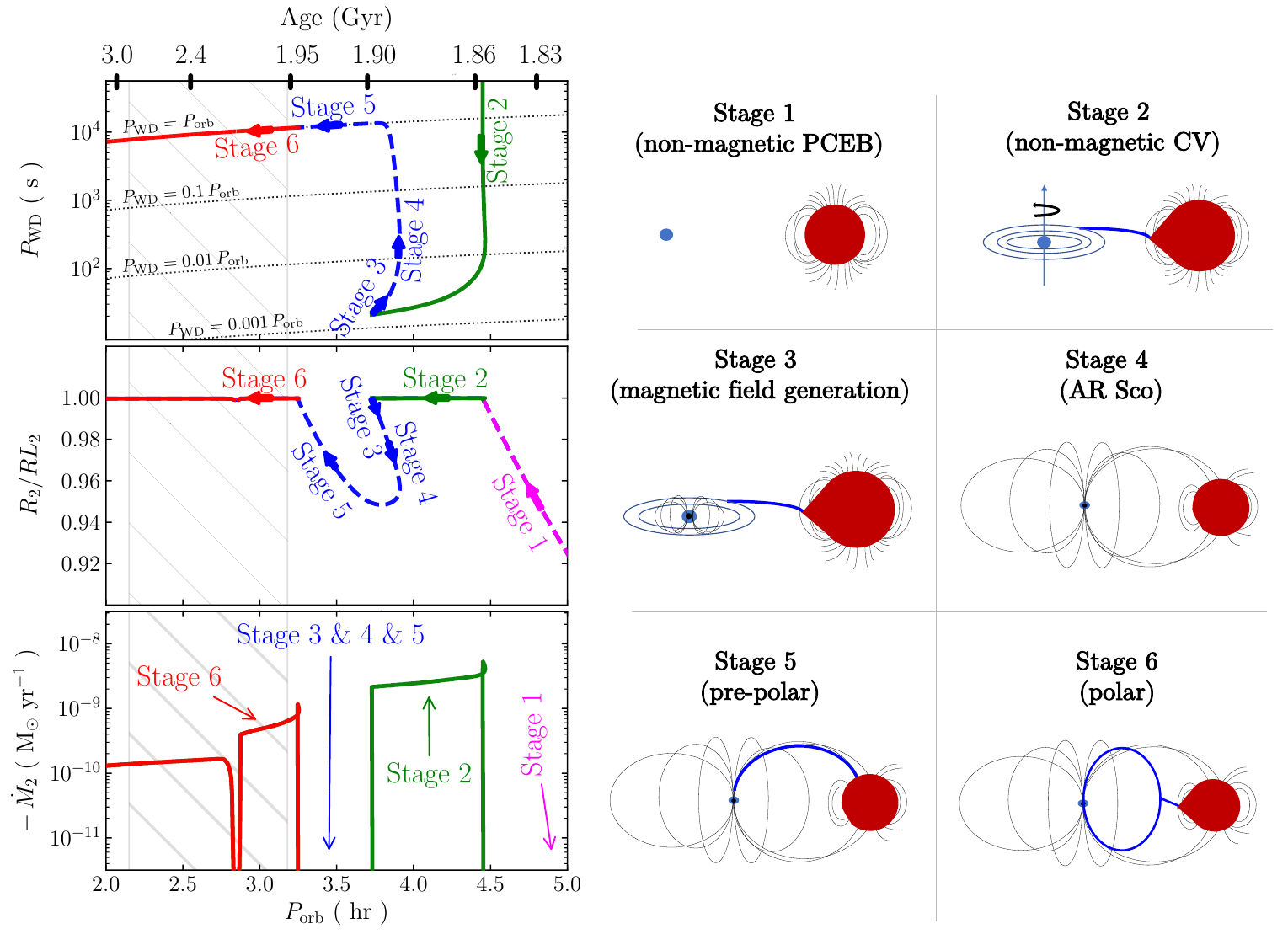}}
\noindent\textbf{Fig.\,1.} Spin period, Roche-lobe filling factor, 
and mass transfer rate as a function of orbital period. The slowly 
rotating white dwarf (stage 1) in a non-magnetic post common envelope binary is spun-up due to accretion (stage 2). In case the white dwarf is crystallizing a strong magnetic field is generated by a dynamo (stage 3) and as soon as it connects with the field 
of the secondary the system becomes detached. 
At first it resembles the radio pulsing white dwarf binary AR\,Sco (stage 4). When the spin of the white dwarf and the orbital motion are synchronised (stage 5) the system becomes a pre-polar before mass transfer starts again (stage 6).  When the secondary becomes fully convective, a second
short detached phase (between $\approx2.07$ and $\approx2.09$\,Gyr) occurs which represents a short version of the classical orbital period gap (hatched region).

\clearpage

\centerline{\includegraphics[width=89mm]{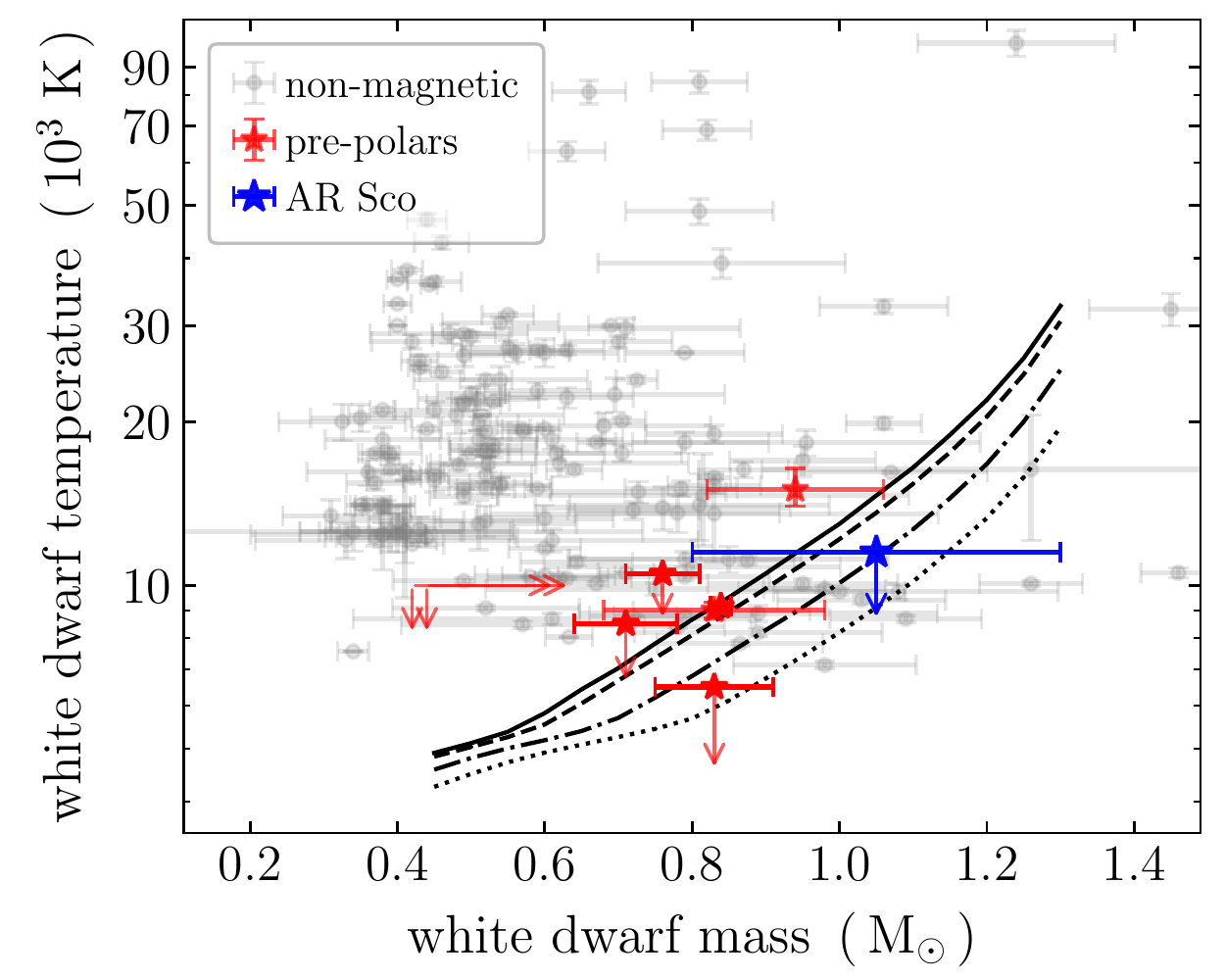}}
\noindent\textbf{Fig.\,2}
Observed white dwarf effective temperatures versus their masses for non-magnetic post common envelope binaries 
from the Sloan Digital Sky Survey (light gray)\cite{nebot-gomezetal11-1,rebassa-mansergasetal12-1}, pre-polars (red)\cite{parsonsetal20-1}, and the first radio pulsing white dwarf binary AR Sco (blue)\cite{marshetal16-1,Garnavich2020}. 
The errors on the observations correspond to 1\,$\sigma$ and arrows indicate upper limits. 
The lines represents the mass fraction of crystallized matter in the white dwarf interiors. Solid, dashed, dash-dotted and dotted lines show the temperature for each white dwarf mass corresponding to the onset of crystallization, 10 per cent of mass is crystallized, 50 per cent and 80 per cent, respectively\cite{bedardetal20-1}. 
As predicted by our scenario, magnetic systems host on average much older and more massive white dwarfs than their non-magnetic counterparts. The trend towards larger masses in magnetic systems is totally consistent with the observed white dwarf mass distribution in cataclysmic variables\cite{zorotovicetal11-1,schreiberetal16-1} which fits with the model prediction that these systems have been CVs in the past. 
In addition, the white dwarfs in AR\,Sco and all
pre-polars are consistent with having crystallizing cores. 

\clearpage

\section*{\large{Methods}}

\subsection{The convective dynamo and scaling laws.}

The magnetic fields of Earth and Jupiter, as well as those of rapidly rotating low-mass stars, are generated by convection-driven dynamos. 
The basic ingredients for these dynamos to work are a strong density stratification, and an extended convection zone. 
To derive the magnetic field strength from fundamental properties of a given planet or star 
the use of scaling laws is required. 
In general it is neither clear which quantities need to be considered in these scaling laws nor 
whether a single scaling law exists for all planets and stars whose magnetic fields are generated by 
the convective dynamo. In the last decades, several scaling laws have been proposed, 
most of them based on the assumption of a 
balance between Coriolis force and Lorentz force\cite{christensen10-1}. 
In recent years a scaling law derived from geo-dynamo models\cite{christensen+aubert06-1} 
which assumes 
the energy flux to be the dominant parameter
has become popular. 
This scaling law has been shown to successfully explain the field strengths of Earth and Jupiter as well as those of rapidly 
rotating fully convective stars$^2$. 

In the first application of the convective dynamo mechanism to white dwarfs 
this energy flux scaling law was assumed$^{20}$. 
The vast majority of white dwarfs consist of Carbon
and Oxygen. 
As a white dwarf cools, the ions in the core begin to freeze in a lattice structure\cite{vanhorn68-1}. 
The phase diagram of the carbon–oxygen mixture is of the spindle form\cite{segretain+chabrier93-1,horowitzetal10-1}, therefore, as soon as the white dwarf has sufficiently cooled and starts to crystallize, 
the solid phase becomes richer in oxygen and sinks while the carbon excess mixes with the outer liquid envelope. 
As a consequence, an oxygen-rich core forms which is significantly denser than the carbon-rich liquid which is redistributed by the Rayleigh–Taylor instability\cite{isernetal97-1,isernetal00-1}. 
Such a strong density stratification in combination 
with convection represent the key ingredients for the dynamo generating the Earth's magnetic fields 
and those of low-mass stars. 
Assuming the same scaling law 
that has been successfully 
applied to the Earth, Jupiter, and rapidly rotating fully convective stars 
for crystallizing white dwarfs predicts field strengths 
of at most a few MG$^{20}$ 
which, if true, would exclude the dynamo from being the mechanism generating the much stronger $1-200$\,MG fields found in white dwarfs in close binary stars. 

Here we argue that this scaling law 
is most likely not applicable to white dwarfs. 
In the magneto hydrodynamic simulations that led to the energy flux scaling law, the magnetic Prandtl number 
($P_{\mathrm{m}}$, defined as the ratio between viscous and magnetic diffusivity) is assumed to be equal to one. However, in both fully convective stars as well as the Earth this number is significantly smaller ($10^{-4}-10^{-5} $ and $10^{-6}$, respectively). 
To quote the discovery paper\cite{christensen+aubert06-1} 
of 
the energy flux scaling law
\emph{``we cannot rule
out an additional dependence on other parameters, in particular the
magnetic Prandtl number. Although the suggested dependence is
weak, it poses a serious problem. Given the large range of extrapolation
over five orders of magnitude from our models to planetary
values of $P_{\mathrm{m}}$, the results obtained from the scaling laws with or without a dependence on $P_{\rm{m}}$ differ substantially.''} 
This early statement has been confirmed more recently by Brandenburg\cite{brandenburg14-1} 
who wrote: \emph{``This leads to a scaling law that has been verified over many orders of magnitude (Christensen et al. 2009). As we have seen, this scaling law must be affected by $P_m$...''}. 
Given that the magnetic Prandtl number for crystallizing white dwarfs has been 
estimated to be $0.58^{20}$, i.e. 
roughly six orders of magnitude larger than that of the Earth and still $4-5$ orders of magnitude larger than that of fully convective low-mass stars, the dynamo could be more efficient in white dwarfs. 

In fact, the comparison between predicted and observed field strengths of single white dwarfs presented by Isern et al.$^{20}$
indeed indicates that observed fields in crystallizing single white dwarfs exceed those predicted by the energy flux scaling law.
The energy flux scaling law is only applicable in the saturated regime of the dynamo, i.e. it only works for rapidly rotating objects with Rossby numbers below $\sim0.1$.
For white dwarfs, this implies spin periods shorter than $\sim90$\,s$^{20}$. 
In white dwarfs that rotate substantially slower, the generated fields should be smaller than those predicted by the 
energy flux scaling law. 
All the observed magnetic single white dwarfs within 20\,pc (Table\,1 of Isern et al.$^{20}$) 
rotate with periods ranging from hours to days.
Despite rotating orders of magnitudes too slow for the energy flux scaling law to be applicable (which is applicable in the saturated regime), the field strengths of these white dwarfs even slightly 
exceed those predicted by the energy flux scaling law. 
This implies that, if a convection driven dynamo works in crystallizing and rotating white dwarfs, 
the energy flux scaling law underestimates the field strengths of slowly rotating single white dwarfs. 
The field strengths of rapidly rotating white dwarfs in the saturated regime should therefore as well 
largely exceed those predicted by the energy flux scaling law. 
A suitably revised scaling law that incorporates the dependence on the magnetic Prandtl number is therefore 
likely to predict fields much stronger than $\sim1$\,MG for white dwarfs that rotate in the saturated regime of the dynamo. 

Based on the above arguments, we here assumed that a rapidly rotating white dwarf ($P_{\mathrm{WD}}\simeq20-30$\,s) with a fractional crystallized mass between 10 and 80 percent can drive a dynamo that generates the large magnetic 
field strengths observed in white dwarfs in cataclysmic variables (up to several 100\,MG). 
We calculate the evolution of close detached white dwarf binaries and cataclysmic variables (CVs) using MESA$^{22}$, 
taking into account the spin-up due to accretion and crystallization due to cooling of the white dwarf 
as well as magnetic field generation if a crystallizing white dwarf rotates rapidly. 

\subsection{Spin-up of the white dwarf.}

If a white dwarf accretes from an accretion disk in a CV its angular momentum and therefore its angular velocity 
increases, while during nova eruptions the white dwarf looses mass and angular momentum.  
According to King et al.$^{23}$, 
the resulting angular momentum balance 
for an accreting white dwarf is given by
\begin{equation}
I\,\frac{{\rm d}\omega}{{\rm d}t} \ = \ 
\alpha \, ( \, -\dot{M}_{2} \, ) \, \left( G \, M_{\rm WD} \, R_{\rm WD} \, \right)^{1/2} \ + \ \left( 1 + \epsilon \right) \left( \dot{M}_{2} \, \eta \, R_{\rm WD}^2 \, \omega \right),
\label{Eq:spinup}
\end{equation}
%
\noindent
where $\omega$ is the white dwarf spin, $I$ its moment of inertia, $G$ the gravitational constant, $\dot{M}_2$ the mass transfer rate averaged over nova cycles, and $M_{\rm WD}$ and $R_{\rm WD}$ are the white dwarf mass and radius.

The first term in the right-hand part of the equation corresponds to the spin-up due to accretion, while the second term represents the spin-down due to material leaving the white dwarf during nova eruptions.
The parameter $0\leq\alpha\leq1$ represents the spin-up efficiency, which might be smaller than one due to mass loss processes not considered in equation Eq.\,\ref{Eq:spinup} (e.g. winds from the inner accretion disk or jets from the boundary layer/white dwarf).
The parameter $0\leq\epsilon=\dot{M}_{\rm WD}/\dot{M}_2\leq1$ 
relates the mass loss of the white dwarf to the mass transfer rate. 
The range of possible values reflects the fact that the white dwarf mass might decrease over a nova cycle, i.e. 
more mass might be expelled during a nova eruption than was 
accreted between two eruptions.
The parameter $0\leq\eta\leq1$ 
represents the square of the radius of gyration of the ejected envelope and depends on the geometry 
of nova eruptions. 

By solving the non-homogeneous differential equation~\ref{Eq:spinup}, the WD spin as a function of the donor mass is given by
\begin{equation}
\omega = \frac{C_1}{C_2} + C_3\,\exp\,{\left(\,C_2\,M_{\rm 2}\,\right)}~,
\end{equation}
%
\noindent
where $M_{\rm 2}$ is the donor mass in solar masses 
and the terms $C_1$, $C_2$ and $C_3$ are given by 
\begin{equation}
C_1 = \alpha\,\frac{A}{k^2}\,\frac{\omega_{\rm bk,0}}{M_{\rm WD,0}}
\end{equation}
\begin{equation}
C_2 = \frac{\eta}{k^2\,M_{\rm WD}} \, \left(\,1+\varepsilon\,\right) + B\,\varepsilon
\end{equation}
\begin{equation}
C_3 =  \left(\,\omega_0-\frac{C_1}{C_2}\,\right)\,\exp{\left(-C_2\,M_{\rm 2,0}\right)}
\end{equation}
%
\noindent
where $\omega_{\rm bk}$ is the break-up white dwarf spin, given by
\begin{equation}
\omega_{\rm bk}
= 
\sqrt{\frac{GM_{\rm WD}}{R_{\rm WD}^3}} = 2\,\pi\,
\sqrt{
  \left(
    \frac{M_{\rm WD}}{{\rm M}_\odot}
  \right)
  \left(
    \frac{R_{\rm WD}}{\rm au}
  \right)^{-3}
}
~~~~{\rm yr}^{-1}~~.
\label{OMEGAWD}
\end{equation}
%
The moment of inertia $I$ is $k^2M_{\rm WD}R_{\rm WD}^2$, with $k$ given by ${0.452+0.0853\log_{10}\left(1 - M_{\rm WD}/1.44 {\rm M}_\odot \right)}$, if ${M_{\rm WD}\lesssim1.368}$~M$_\odot$, and $\approx0.275$, if ${M_{\rm WD}\gtrsim1.368}$~M$_\odot$. 
The constants $A$ and $B$ are given by
\begin{equation}
A = \left[ \frac{1-0.61\,\left( \, M_{\rm WD,0}/{\rm M}_\odot \, \right)^{4/3}}{1-0.61\,\left( \, M_{\rm WD}/{\rm M}_\odot \, \right)^{4/3}}  \right]^{3/4} ~~,
\end{equation}
\begin{equation}
B = \frac{3.05\,\left( \, M_{\rm WD}/{\rm M}_\odot \, \right)^{4/3} - 1}{3\,\left[ 1 - 0.61\left( \, M_{\rm WD}/{\rm M}_\odot \, \right)^{4/3} \right]}  \ + \ \frac{0.051\,\left( \, M_{\rm WD}/{\rm M}_\odot \, \right)}{k\,\left[ \, 1 - 0.69 \left( \, M_{\rm WD}/{\rm M}_\odot \, \right) \right]}~~.
\end{equation}
The sub-index $0$ in some quantities indicates their initial values just after the common-envelope evolution.

\subsection{Synchronisation of the white dwarf.}

In accreting high-field magnetic white dwarf binaries
the magnetic field dominates the accretion process. 
For sufficiently strong white dwarf magnetic moments, the
overflowing material is channelled along the white dwarf's magnetic field lines as soon as the magnetic pressure dominates the gas ram pressure.
For such strong white dwarf magnetic fields, the interaction of the magnetospheres of the white dwarf and the donor cause
a synchronizing torque on the white dwarf\cite{campbell85-1}.

We assume here that the onset of synchronization occurs when the synchronizing torque is greater than the accretion torque (assuming that the magnetospheres of the white dwarf and the donor are connected)$^{26}$, i.e. 
\begin{equation} 
\begingroup
{\textstyle
\frac{\mu_{\rm WD}}{10^{33}\,{\rm G\,cm^3}} >  1.11 \, \left( \frac{-\dot{M}_2}{10^{16} \,{\rm g/s}}  \right) \, \left( \frac{M_{\rm tot}}{{\rm M}_\odot} \right)^{1/3} \, \left( \frac{M_2}{{\rm M}_\odot} \right)^{-1} \, \left( \frac{P_{\rm orb}}{{\rm hr}} \right)^{1/3} \, \left( \frac{B_{\rm 2}}{100\,{\rm G}} \right)^{-1} \, \left[ 0.5 - 0.277\,\log_{10}\,\left(\frac{M_2}{M_{\rm WD}} \right)\right]^{2} 
}
\endgroup
\label{Eq:sync}
\end{equation}
%
\noindent
where $\mu_{\rm WD}$ is the white dwarf magnetic moment, $M_2$ is the donor mass, $M_{\rm tot}=M_{\rm WD} + M_2$ is the binary total mass, $P_{\rm orb}$ is the orbital period and $B_2$ is the strength of the magnetic field of the secondary star.  

Once the condition provided by Equation~\ref{Eq:sync} is satisfied, the spin-down process starts.
For simplicity, we assume that the white dwarf spin decreases exponentially on a given time-scale, i.e.
\begin{equation} 
\omega(t)= \Omega + C e^{-t/\tau}
\label{Eq:spindown}
\end{equation}
\noindent
where $\tau$ is the synchronisation time-scale, $\Omega$ the orbital angular velocity, $\omega$ the WD spin, and $C=\omega_0-\Omega$ is a constant (because variations in $\Omega$ are negligible during the spin-down phase). 
This assumption gives 
\begin{equation}
    \frac{{\rm d}\omega}{{\rm d}t}=  \frac{(\Omega - \omega)}{\tau}.
\end{equation}
This equation of course represents a very crude approximation of a very complicated process. 
The torque on the primary as a function of asynchronism has been calculated previously in great detail\cite{campbell10-1}, but only for rather small asynchronism. 
Performing such detailed calculations of the magnetic field interaction for the large asynchronism in our systems, 
is beyond the scope of this paper.
For the purpose of this paper, we assume that the total angular momentum loss by the white dwarf is transferred to the secondary, incorporated in the orbital motion, and that this occurs on the time-scale $\tau$, i.e. 
\begin{equation}
  \dot{J}_{\rm orb}= - \frac{\rm d}{{\rm d}t}(\omega I) = I \frac{(\omega - \Omega)}{\tau} > 0. 
\end{equation}
In this simple model the torque is always decreasing with increasing synchronism which might not be correct\cite{campbell10-1}. However, 
this should not affect the validity of our model as 
only the amount of spin angular momentum transferred to the orbit and the time scale on which this happen are important 
for the secular evolution of a given system. 

\subsection{Reduced magnetic braking.}

For secondary stars still containing a radiative core, 
angular momentum loss through magnetic wind braking
drives mass transfer in CVs$^{6,11}$.
The strength of magnetic braking depends significantly on 
the extension of 
wind zones compared to 
dead zones. The latter correspond to 
regions where gas is prevented from 
escaping the star due to the 
pressure of closed magnetic 
field loops, i.e. the magnetic 
pressure is greater than the thermal
pressure\cite{Mestel_1968}. 
If the white dwarf in a CV is strongly magnetic and the magnetospheres of both stars connect,
the wind zone is reduced which 
leads to a decrease of orbital angular momentum loss due 
to magnetic braking\cite{WW02,LWW94}.  
For short orbital periods and/or strong magnetic fields, angular momentum loss through magnetic braking can be completely suppressed. 

The reduction of magnetic braking for a given $\mu_{\rm WD}$ can be parametrized 
with the fraction of open field lines $(\Phi)$.
In the extreme case of no wind zone, i.e. $\mu_{\rm WD}$ is strong enough to prevent any magnetic braking, 
${\Phi=0.0}$. 
On the other hand, in case the impact of the magnetic moment is negligible, the second dead zone does not exist and magnetic braking is not reduced, leading to ${\Phi=0.258}$.
This prescription for the reduction of MB 
depends on the binary parameters 
and $\mu_{\rm WD}$, and therefore allows us to incorporate the impact of strong WD magnetic fields in binary evolution codes\cite{WW02}$^{,25}$. 

The resulting angular momentum loss due to magnetic braking in the presence of a strongly magnetic white dwarf $(\dot{J}_{\rm MB,mag})$ can be written as 
\begin{equation}
\dot{J}_{\rm MB,mag} \ = \ 
\dot{J}_{\rm MB,non-mag} \, \left( \frac{\Phi}{0.258} \right)^{5/3} \, ,
\label{EQN1}
\end{equation}
\noindent
where 
$\dot{J}_{\rm MB,non-mag}$ is the angular momentum loss due to magnetic braking in non-magnetic CVs.
In this work we used the widely adopted RVJ prescription\cite{RVJ}, with $\gamma=3$, i.e.
\begin{eqnarray}
\dot{J}_{\rm MB,non-mag} \ = \ -3.8\times 10^{-30}\,\left(\frac{M_2}{\rm g}\right)\,R_\odot^4\,\left(\frac{R_2}{{\rm R}_\odot}\right)^3\,\left(\frac{\Omega_2}{\rm s^{-1}}\right)^3,
\label{Eq:MB-RVJ}
\end{eqnarray}
%
\noindent
where $R_2$ is the donor radius and $\Omega_2$ is the donor spin, which is the same as the orbital angular velocity.
This reduced magnetic braking model has been incorporated in binary population models of magnetic CVs and has been shown to convincingly reproduce the orbital period distribution of magnetic CVs$^{25}$ 
and we here implemented the same prescription in MESA.  

\subsection{Modelling magnetic CV evolution with MESA.}

The above described effects were incorporated in the stellar evolution code 
Modules for Experiments in Stellar Astrophysics (MESA) (version r10108)\cite{Paxton2011, Paxton2013, Paxton2015, Paxton2018}$^{,22}$. 
The MESA equation of state is a blend of the OPAL\cite{Rogers2002}, SCVH\cite{Saumon1995}, PTEH\cite{Pols1995}, HELM\cite{Timmes2000}, and PC\cite{Potekhin2010} equation of states. Radiative opacities are primarily from OPAL\cite{Iglesias1993,Iglesias1996}, being the regime low-temperature dominated\cite{Ferguson2005} or high-temperature, Compton-scattering dominated\cite{Buchler1976}.
Electron conduction opacities are also included\cite{Cassisi2007}. Nuclear reaction rates are a combination of rates from NACRE\cite{Angulo1999}, JINA REACLIB\cite{Cyburt2010}, plus additional tabulated weak reaction rates\cite{Fuller1985, Oda1994,Langanke2000}.
Screening is included\cite{Chugunov2007} as well as thermal neutrino loss\cite{Itoh1996}. 
Roche lobe radii in binary systems are computed using Eggleton's fit\cite{Eggleton1983} and mass transfer rates in Roche lobe overflowing binary systems are determined following Ritter's prescription\cite{Ritter1988}.

The parameters of our standard model were set as follows. 
During the whole CV evolution we assumed that all material accreted by the WD is lost through nova eruptions. In Equation~\ref{Eq:spinup}, this corresponds to setting $\epsilon=0$.
For the geometry of nova shells, we assumed that mass leaves the WD with some asymmetry (i.e. neither completely spherical nor radial), which implies values of $\eta$ in Eq.~\ref{Eq:spinup} between $2/3$ and $1$. We set ${\eta=2.5/3}$.
Additional mass loss from the system\cite{matthews15-1,coppejans20-1} was taken into account by 
using $\alpha=0.75$ in Equation~\ref{Eq:spinup}. 
The donor magnetic field was assumed to be ${B_2=1}$~kG, which is consistent with recent measurements of field strengths of low-mass stars\cite{reiners07-1}.
We fixed the WD mass to $0.8$~M$_\odot$, which corresponds to the average value found among observed CVs$^{29,}$\cite{mcallister19-1}.
For the white dwarf magnetic field, we assume ${B_{\rm WD}=60}$~MG, which is slightly stronger than the average field strength of polars$^{7}$, 
but corresponds to the average field strength found in pre-polars$^{9}$. 
We assumed that the dynamo is efficient only if the core of the white dwarf is crystallizing (between 10 and 80 per cent crystallized) and the spin period of the white dwarf is $\simeq20-30$\,s. %
Finally, for the synchronisation time-scale, we set ${\tau=1}$~Myr, which is consistent with the expected rapid spin-down in close binaries harbouring a rapidly rotating magnetic white dwarf such as AR~Sco$^3$. 

Using this set of parameters our model can reproduce the observations of magnetic detached and semi-detached close white dwarf binary stars (Supplementary Figure\,1). 
According to this model, pre-polars and AR\,Sco have been CVs in the past and at that time the strong magnetic fields of their white dwarfs were generated. This prediction is in perfect agreement with the 
cool white dwarf temperatures of these detached systems, with them being close to Roche-lobe filling, and with their preferred occurrence in the 3-5\,hr orbital period range. 

We emphasise that no fine-tuning was required to obtain this agreement. In contrast, the model naturally explains the observations and variations of the key parameters do not affect the general predictions (Supplementary Figures\,2, 3 and 4).
For instance, by changing the spin-up parameters (i.e. $\alpha$, $\eta$ and $\epsilon$), the only effect is to accelerate or delay the spin-up of the white dwarf and slightly change the minimum spin period that can be reached$^{23}$. 
For all reasonable values of these parameters the white dwarf will manage to spin-up to sufficiently large spin rates 
so that the conditions for generating strong magnetic fields can eventually be met.

Similarly, varying the synchronisation time-scale $\tau$ affects the synchronising torque but does not change the evolutionary sequence (Supplementary Figure\,2). 
The shorter $\tau$, the stronger the spin-down torque and therefore the longer the detached phase. 
Only for $\tau\gtrsim100$~Myr, the detached phase can be fully 
prevented if, in addition, the magnetic field strength of the white dwarf does not exceed $\sim200$\,MG. 
The latter condition arises as for very strong fields reduced magnetic braking alone can cause the system to detach in full analogy to the mechanism producing the classic orbital period gap. 

A slightly stronger constraint on the synchronisation time scale results from
the fact that it should not exceed the duration of the generated detached phase as otherwise the observed synchronised pre-polars would remain unexplained. 
For white dwarfs rotating near break-up, in order to have a synchronised system after the detached phase, time scales cannot exceed $30-40$\,Myr.
This maximum time scale is definitely not a problem for our model, since the synchronisation time scales for the range of magnetic field strengths we consider here, i.e. those of polars, pre-polars and AR~Sco, are expected to be well below $30-40$\,Myr\cite{stiller18-1}. 

The overall evolution does also not sensitively depend on the assumed white dwarf and secondary star magnetic fields. 
Varying these field strengths changes only slightly the condition for synchronisation (Equation~\ref{Eq:sync}). 
For weaker/stronger $B_2$, this condition becomes more/less restrictive. 
Increasing or decreasing $B_{\rm WD}$ affects somewhat the orbital period at which synchronism can be obtained, which is longer for stronger fields (Supplementary Figure\,3).
The white dwarf magnetic field strengths also affects the duration of the detached phase. The stronger the white dwarf magnetic field, the more angular momentum loss due to magnetic braking is reduced and the longer it takes the system to evolve back into a semi-detached configuration.

The largest and most critical assumption in our model certainly
are the conditions we assume for magnetic field generation which represent a strong but reasonable simplification. 
According to the most popular scaling law, 
the dynamo saturates for fast rotation (in white dwarfs 
for spin periods below 90\,s$^{20}$) 
and the produced field strength depends then solely on the 
convective energy flux. However, as outlined above, 
the efficiency 
of the dynamo 
is also expected to significantly depend on the magnetic Prandtl number which is not considered in any 
currently available scaling law.   
As long as a reliable scaling law taking into account this dependency 
is not available, describing the evolution of close magnetic white dwarf binaries
needs to be based on 
simple assumptions such as those made in this paper. 
As soon as realistic scaling laws for the dynamo operating in crystallizing white dwarfs are available, we plan to incorporate those together with detailed white dwarf 
models into our simulations. 

The aim of this paper is to show how the evolution of close white dwarf binaries is affected if, and that is a reasonable assumption,
crystallization and fast rotation can generate a strong magnetic field. We would expect the generated field strength in these models to depend on the rotation rate and the energy flux in the convection zone. 
Given the large range of orbital periods at the end of common envelope evolution, and the therefore large variety of evolutionary time scales towards the CV phase, we expect that such a more detailed model can explain the large range of field strength observed in magnetic CVs including those of intermediate polars. In case of the latter the field is simply not strong enough to synchronise the white dwarf spin with the orbital motion and the system does not enter a detached phase apart from the usual $2-3$\,hr orbital period gap. 

In the Supplementary Information we present a detailed discussion of different evolutionary channels predicted by our model and estimate the fraction of magnetic and non-magnetic white dwarfs in CVs taking into account important parameters for binary star evolution\cite{RVJ,zorotovicetal10-1,zorotovic14-1,bellonietal18-1,MD17}
as well as
potential accretion heating and chemical impurities of the core of the white dwarf\cite{townsley+bildsten04-1,epelstainetal07-1,isernetal98-1}.
We also elaborate on the implications of 
our scenario for the generation of magnetic fields in hot single white dwarfs\cite{charpinetetal09-1} as well as magnetic white dwarfs in wide binaries\cite{landstreet+bagnulo20-1}. We find that these magnetic fields are most likely generated by alternative mechanisms such as fossil fields resulting from main sequence star mergers\cite{Ferrario2009,schneideretal19-1}, double white dwarf mergers\cite{garcia-berroetal12-1}, or common envelope mergers\cite{toutetal08-1,Wickramasinghe2014,Briggs2015}.

\clearpage

\begin{addendum}

\item[Correspondence] Correspondence and requests for materials should be addressed to M.R.S. \\ (email: matthias.schreiber@usm.cl).

\item We thank the referee Chris Tout and two anonymous referees for their helpful comments.
M.R.S. acknowledges support from the Millennium Nucleus for Planet Formation (NPF) and Fondecyt (grant 1181404). D.B. was supported by the grant {\#2017/14289-3}, S\~ao Paulo Research Foundation (FAPESP) and ESO/Gobierno de Chile. B.T.G. was supported by the UK STFC grant ST/P000495. S.G.P acknowledges the support of a STFC Ernest Rutherford Fellowship. M.Z. acknowledges support from CONICYT PAI (Concurso Nacional de Inserci\'on en la Academia 2017, Folio 79170121) and CONICYT/FONDECYT (Programa de Iniciaci\'on, Folio 11170559)

\item[Author contributions] 
All authors contributed to the discussion and writing of this article. 
M.R.S. and D.B. developed the idea and carried out the simulations. 
S.G.P. and B.T.G provided an observational overview of magnetic white dwarfs and recent observations of pre-polars. 
M.Z. provided crystallization temperatures for different white dwarf masses based on published white dwarf evolutionary sequences and estimated relative numbers of magnetic and non-magnetic CVs using binary population models.

\item[Competing Interests] The authors declare that they have no competing interests.

\item[Data availability] The data presented in this work is available upon request. In addition, the \textsc{MESA} files required to reproduce our simulations will be publicly available at the \textsc{MESA} Zenodo community (https://zenodo.org/communities/mesa/).

\item[Code availability] \textsc{MESA} is publicly available (http://mesa.sourceforge.net/).

\end{addendum}


\section*{\large{Supplementary Discussion and Figures}}

\subsection{The fraction of magnetic white dwarfs in CVs.}

We have shown that a crystallization and rotation driven dynamo can explain the properties 
of all close and strongly magnetic white dwarf binaries and 
the absence of 
magnetic white dwarfs in young detached white dwarf binary stars. In what follows 
we discuss under which conditions our model is likely to also reproduce the observed fraction of strongly magnetic white dwarfs in CVs.  
The measured fraction of strongly magnetic white dwarfs in CVs is more than one third (15 out of 42 systems within 150\,pc$^1$). If the briefly detached pre-polars and AR\,Sco are included in this sample 
(the four pre-polars SDSS\,0303+0054, WX\,LMi, SDSS\,J1212+0136, SDSS\,J1452+2045 and AR\,Sco are within 150\,pc$^{3,10}$), the fraction of systems with magnetic white dwarfs further increases to 20 of 47 systems corresponding to $43\pm6$ per cent. 

At first glance, our model might seem to produce too many magnetic white dwarfs in CVs as every white dwarf in a non-magnetic CV accretes angular momentum and is therefore expected to become rapidly rotating in a short period of time. In addition, if the white dwarf continuously cools it should start to crystallize and one might expect that our model predicts every white dwarf in a CV to eventually develop a strong magnetic field. However, accretion can increase the core temperature of white dwarfs in CVs$^{77}$ and prevent or delay crystallization.
In addition, the generation of a strong magnetic field through the rotational and crystallization driven dynamo 
only works if fast rotation and a crystallizing core with an outer convection zone are present 
at the same time. The strength of the magnetic field generated by this dynamo
depends on the convective energy flux$^2$ and most likely also on the magnetic Prantdl number. 
However, in the absence of a scaling relation including the dependence on the latter, 
it remains unclear at which stages of crystallization the conditions 
are suitable for generating strong magnetic fields. In our model we assumed that in addition to fast rotation at least 10 per cent and not more than 80 per cent of the core need to be crystallized for the dynamo to be able to generate a strong magnetic field but this assumption is rather arbitrary. In what follows we describe the 
three different types of CVs that are non-magnetic according to our model and explain how the critical core mass ratios and other important parameters impact the predicted relative numbers of magnetic and non-magnetic CVs.

First (channel 1), our model predicts the white dwarfs in CVs that did not yet accrete enough angular momentum remain non-magnetic. How long this phase takes,
and therefore also the relative number of CVs that are expected to be found in this stage in observed samples, depends on the efficiency of the accretion of angular momentum and the mass transfer rate (see equation\,1 in the main text) which in turn depends on the angular momentum loss. 
For instance, CVs born with fully convective secondaries are expected to be driven by gravitational radiation only and typically have mass transfer rates 1-2 orders of magnitude lower than systems with earlier secondary stars. 
Therefore, the white dwarfs born with fully convective secondary stars will take longer to spin up to near-break velocities than those with earlier secondary stars that start mass transfer at longer orbital periods. However, as the spin up time scale is in general short compared to the secular time scale of CVs, we expect binary population synthesis based on our model to predict rather few CVs to be in this non-magnetic CV state.

Second (channel 2), CVs with white dwarfs that did not yet start the crystallization process in their cores remain non-magnetic CVs until the required condition is reached. How long this takes 
determines how many of such systems are predicted to exist in observed samples and depends on the fraction of crystallized core-mass that is assumed to be necessary to make the dynamo work (we assumed 10 per cent), on the time it takes the system to evolve from the common envelope to the CV phase, and whether accretion can heat up the core$^{77,78}$ to avoid or delay crystallization. 
Assuming core heating is inefficient, according to recent white dwarf models$^{24}$ the time needed for a 0.6, 0.8 and 1.0~\Msun white dwarf to have 10 per cent of its interior crystallized is approximately 3.2, 1.9 and 1.1~Gyr. This time span might be substantially increased for smaller white dwarf masses as the magnetic field needs to diffuse through a radiative zone which sits on top of the convective mantle. Only for massive white dwarfs the convective mantle extends close to the surface$^{20}$. 
This implies that detached binaries that emerged from common envelope evolution with short orbital periods, i.e. close to the start of mass transfer, can remain non-magnetic CVs for $\gappr\,3$\,Gyr. If core heating due to accretion is efficient and prevents the white dwarf from crystallizing during the CV phase, these CVs will even remain non-magnetic forever. Therefore a non-negligible number of non-magnetic CVs descending from short period post common envelope binaries is predicted to exist in observed samples of CVs.  

Finally (channel 3), CVs containing white dwarfs that significantly crystallized before being spun up to fast rotation due to accretion, will not develop a strong magnetic field. In other words, if the white dwarf had enough time too cool so that a large fraction of its core is crystallized before the secondary fills its Roche-lobe,
the convective energy flux in the outer core will not be sufficient to generate a strong magnetic field. 
These systems will remain non-magnetic forever under the assumption that core heating due to accretion is inefficient. Even if accretion heating can melt a significant fraction of the core, this process is supposed to take several Gyr$^{78}$, i.e. the system would remain non-magnetic for a large fraction of its CV lifetime.   
Thus, whether a given system becomes a non-magnetic CV formed through this channel depends mostly on the time it spends between the white dwarf formation, i.e. the common envelope phase, and the onset of accretion. How many systems of a predicted CV population spend enough time in this detached white dwarf binary phase depends sensitively on the crystallized mass fraction (we assumed 80 per cent) at which the dynamo becomes unable to generate a strong magnetic field, on the inter-correlated initial orbital period, mass ratio, and eccentricity distributions of the progenitor systems, on the common envelope efficiency, and on the strength of angular momentum loss due to magnetic braking. 

While the relative number of non-magnetic CVs predicted by our model through channel 1 is most likely very small, we expect that the other two groups 
of systems can easily explain the $\sim\,50-70\,$ per cent of 
non-magnetic systems in the observed CV sample. Which of the two scenarios contributes more to the fraction of non-magnetic CVs depends on several assumptions. 
For example, for a large common envelope efficiency,  
relatively inefficient magnetic braking, and a critical crystallized mass fraction below 80 per cent, 
the majority of systems will evolve through channel 3 and will remain non-magnetic for at least several Gyr. 
In contrast, for a small common envelope efficiency, efficient magnetic braking, a large value for the critical fraction of crystallized core mass, and inefficient core heating, 
most CVs will eventually become magnetic but many of them 
may spend several Gyr as non-magnetic systems before the dynamo starts operating (channel 2).

The parameters (strengths of magnetic braking, critical crystallized mass fractions, common envelope efficiency, efficiency of core heating) are essential for predicting the evolution of close white dwarf binaries 
and for determining the fraction of strongly magnetic CVs predicted by our model. Unfortunately, several of these parameters are currently neither theoretically nor observationally well constrained. 
The common envelope efficiency is most likely relatively small as indicated by observational surveys of detached white dwarf binaries$^{73}$ and heating of the core is likely to play an important role$^{77,78}$.
In contrast, the strength of magnetic braking is unknown$^{6}$
and the same is true for the critical 
mass fractions (as detailed simulations for the dynamo we propose here do not yet exist).
In addition, the initial binary parameter distributions usually assumed in CV population models are uncorrelated
which is inconsistent with observational surveys performed in the past decades$^{76}$. These surveys show that the orbital period, primary mass, mass ratio and eccentricity distributions are inter-correlated. This is important because the combination of these parameters determines whether a given binary system will survive or not the common-envelope phase and which of those that survive and become CVs within the Hubble time, have relatively long or short detached lifetimes (possibly becoming slightly or highly crystallized) before the onset of mass transfer. 

Based on binary population models we performed 
in the past$^{74,75}$, 
using standard parameters for CV evolution (common envelope efficiency of 0.25, magnetic braking according to Rappaport et al.$^{44}$, assuming the critical crystallized mass fractions used in this paper (i.e. 10 and 80 per cent), and ignoring core heating of the white dwarf through accretion, we estimate the predicted fraction of magnetic systems to be roughly 
50 per cent. While this value slightly exceeds the currently observed number of magnetic CVs, relatively small changes in the prescription for magnetic braking and/or the 
critical mass fractions of crystallized matter in the core of the white dwarf can easily bring into agreement the observed fractions and the predictions of our model. 
For the same assumptions but including efficient core heating, the predicted fraction of magnetic CVs decreases to $\sim\,20$ per cent. 
This estimate is below the measured fraction but our conclusion remains unchanged as relatively small changes in uncertain parameters such as critical core mass and magnetic braking could increase the number of magnetic systems to match the observed 
fraction. 

For the sake of completeness, as an additional unknown parameter, we emphasize that also the metallicity of the white dwarf progenitor star could play a role since the settling of neutronized impurities like 22Ne and 56Fe caused by solidification can release more energy than the separation of the C/O mixture which significantly affects the cooling of the white dwarf and the crystallization process$^{79}$.  
 
As a final remark, we state that the small number of observationally discovered pre-polars and the even smaller number of AR Sco-like systems are in agreement with the relatively short time scale that our model predicts for these systems.

\subsection{Implications for single white dwarfs and white dwarfs in wide binaries.}

The model presented in this paper describes a new evolutionary sequence for close white dwarf binaries which explains the origin and the relative numbers of strongly magnetic white dwarfs in these systems. While solving the puzzle for the occurrence of strongly magnetic white dwarfs in close binaries represents a key result on its own, we here briefly mention the implications for 
single white dwarfs and white dwarfs in wide binary stars. 

It is clear that the dynamo proposed here to operate in CVs cannot explain all strongly magnetic single white dwarfs because it requires fast rotation and crystallization. A non-negligible number of strongly magnetic single white dwarfs$^{7}$ are hot and therefore most likely not in the process of crystallization and, even worse, several of the highest field single white dwarfs are known to have rather long spin periods$^7,80$. While field strengths of the order of $0.1-1$\,MG might be generated in crystallizing single white dwarfs with rotation periods of the order of hours or days, strong magnetic fields in single white dwarfs must be generated at least partly by a different mechanism. 
The same conclusion holds for the recently discovered strongly magnetic white dwarfs in wide binaries$^{81}$. In most cases the separation between the two stars in these binaries exceeds several hundred astronomical units and therefore a previous phase of mass transfer and spin-up of the white dwarf can be excluded. 

Similarly, the main mechanism for generating these strong magnetic fields in young single white dwarfs and wide binary stars must not be operating 
in close binaries as we would otherwise observe strongly magnetic white dwarfs in young detached close white dwarf binaries. 
The candidate scenarios that may fulfill this conditions are 
white dwarf mergers$^{19}$, fossil fields$^{17}$,
or the common envelope merger scenario$^{18,84,85}$.
Given that the most likely progenitors of magnetic white dwarfs according to the fossil field scenario are main sequence star mergers$^{82,83}$, all three scenarios require single magnetic white dwarfs to form from binary stars. 
This implies that all scenarios require initial triple star configurations to produce strongly magnetic white dwarfs in binaries. 
As triples are intrinsically less common than binary stars, all three models may produce more strongly magnetic single white dwarfs 
than strongly magnetic white dwarfs in close and wide binaries. 
Which of the mechanisms is most suitable to explain magnetic single white dwarfs and magnetic white dwarfs in wide binaries depends on the relatively poorly constraint distribution of initial triple star configurations.

We conclude that the crystallization and rotation driven dynamo proposed here explains the large number of magnetic white dwarfs in CVs, as well as the existence and characteristics of AR\,Sco and the pre-polars, and might be complemented by e.g. the double white dwarf merger and/or the common envelope merger scenario which are good candidates for explaining strongly magnetic single white dwarfs.


\centerline{\includegraphics[width=90mm]{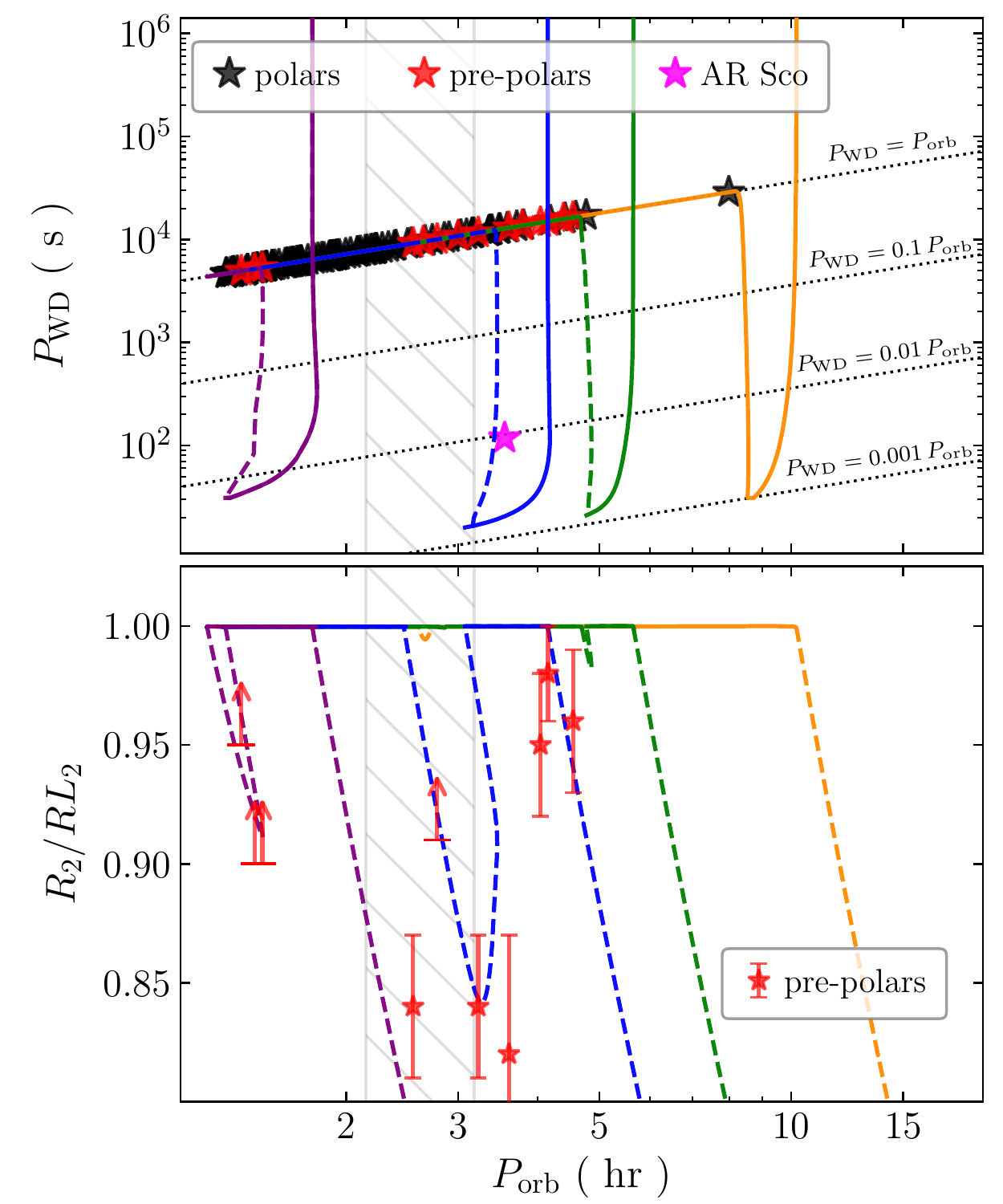}}
\noindent\textbf{Supplementary Figure\,1.} Comparison of different evolutionary 
tracks with observations. 
The detached phases (dashed lines) explain the existence of only cool magnetic white dwarfs in detached binaries that are close to Roche-lobe filling because the magnetic field was generated during the CV phase when the white dwarf was old and crystallizing. 
The model also offers an explanation for AR\,Sco which in fact is a progenitor of 
the pre-polars. In AR\,Sco, nicely reproduced by the blue track, the secondary detached from its Roche-lobe but spin and orbital motion are not yet synchronised. The subsequent detached synchronised phase explains the orbital periods, Roche-lobe filling factors, and white dwarf temperatures of pre-polars. 
To generate the tracks we only changed the initial orbital period ($P_{\mathrm{i}}$) and initial secondary mass ($M_2$), as well as the white dwarf spin period ($P_{\mathrm{WD}}$) when the magnetic field is generated. The parameters we used were the following
$P_{\mathrm{i}}=0.14$\,days, $M_2 = 0.17$\,\Msun, $P_{\mathrm{WD}}=30$\,s (purple),
$P_{\mathrm{i}}=0.68$\,days, $M_2=0.55$\,\Msun, $P_{\mathrm{WD}}=15$\,s (blue),
$P_{\mathrm{i}}=0.925$\,days, $M_2=0.75$\,\Msun, $P_{\mathrm{WD}}=20$\,s (green),
$P_{\mathrm{i}}=1.544$\,days, $M_2=1.2$\,\Msun, $P_{\mathrm{WD}}=30$\,s (orange). In all tracks the white dwarf mass was assumed to be $0.8$\,\Msun$^{29}$
and the generated magnetic field strengths was assumed to be $60$\,MG. 
The errorbars on the observations correspond to 1$\sigma$.

\clearpage

\centerline{\includegraphics[width=90mm]{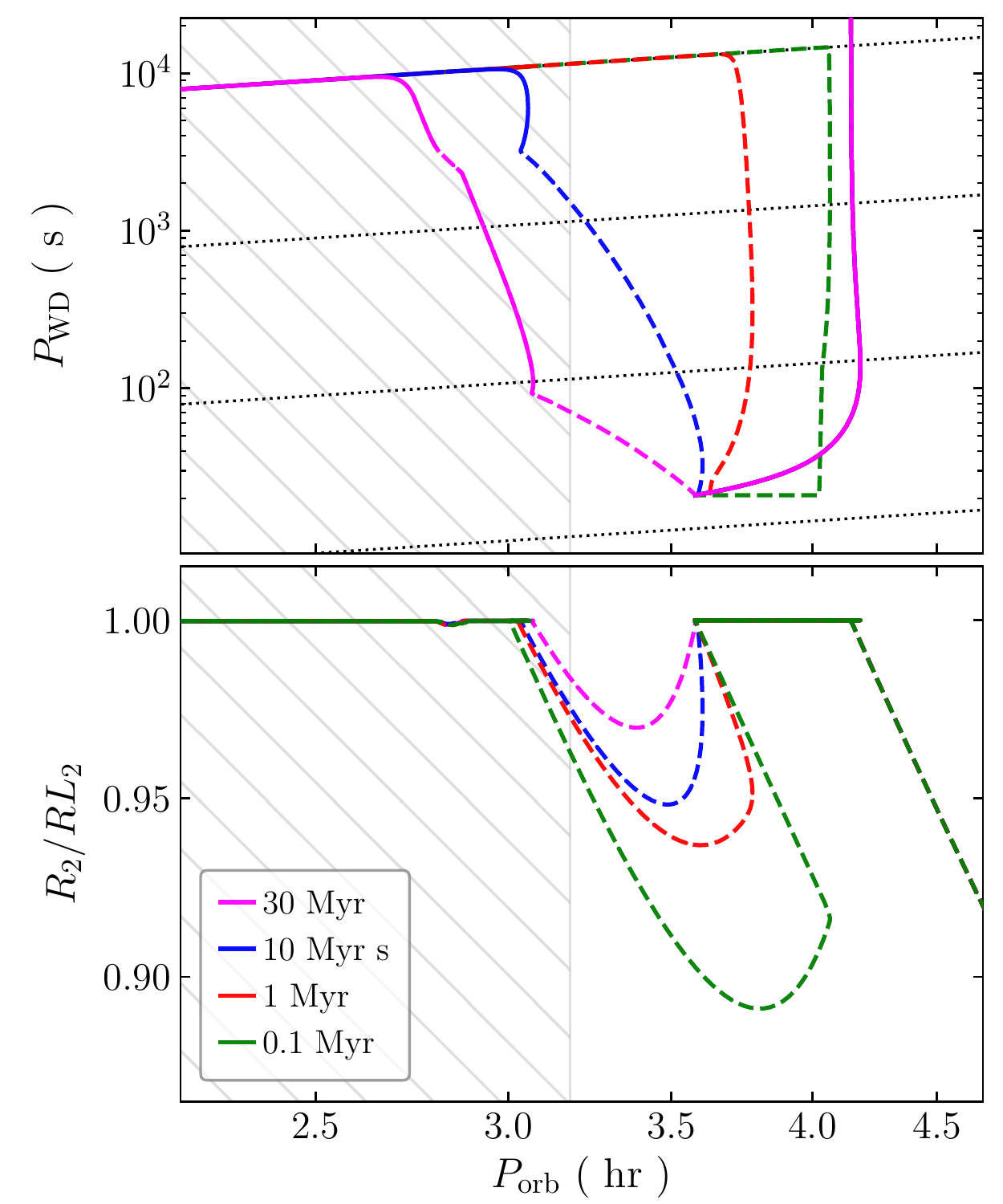}}
\noindent\textbf{Supplementary Figure\,2.} Evolutionary sequences highlighting the impact of different choices of  the synchronisation time scale.
The shorter the time scale, the more pronounced and the longer the detached phase. 
For the longest time scales (blue, purple), accretion starts before the system synchronised. 
Time scales exceeding $1$\,Myr are most likely not realistic.

\clearpage

\centerline{ \includegraphics[width=90mm]{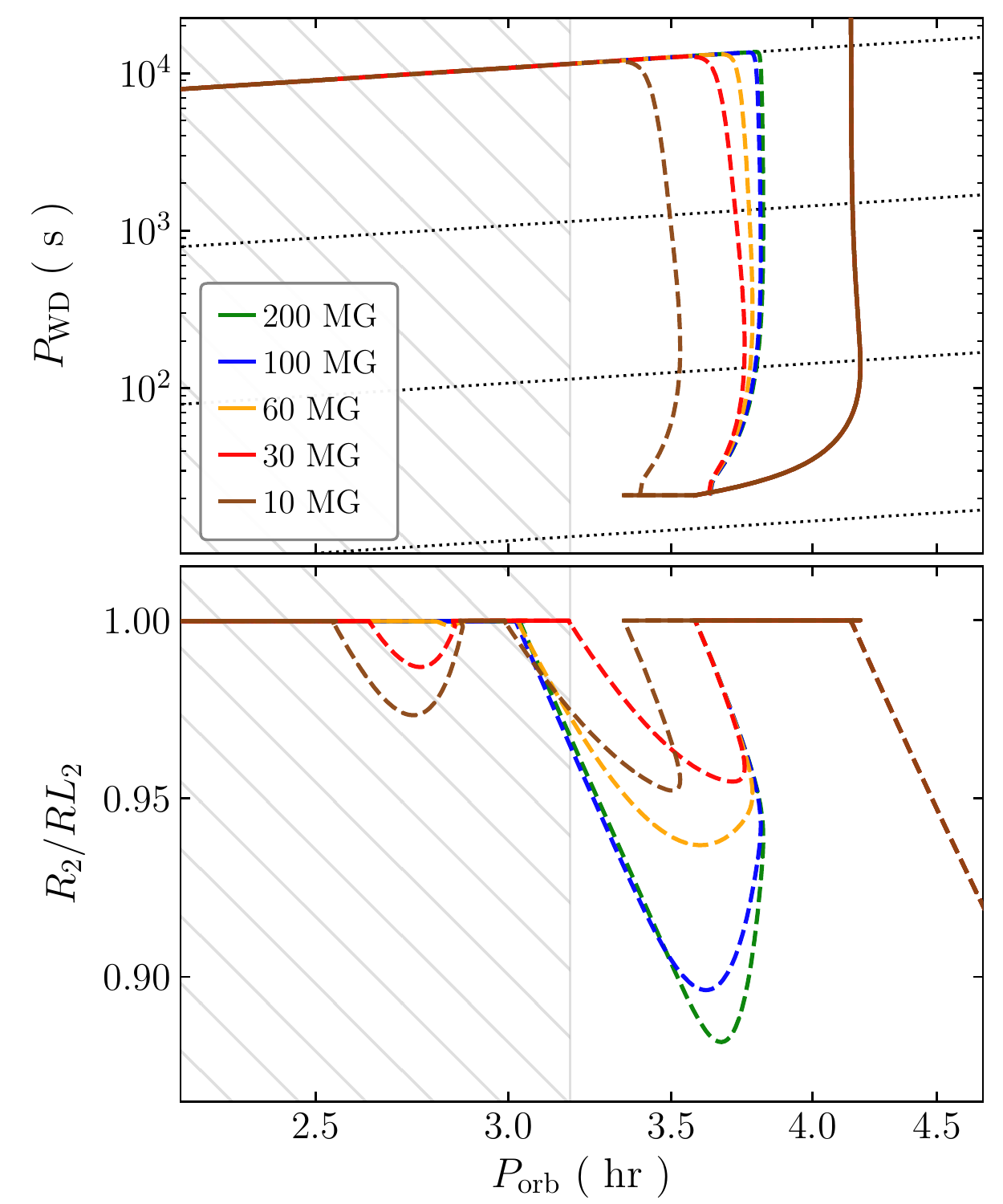}}
\noindent\textbf{Supplementary Figure\,3.} Evolutionary sequences highlighting the impact of the white dwarf magnetic field strength. 
The stronger the generated field, the longer the detached 
phase as angular momentum loss due to
magnetic braking is more reduced. For the weakest field strengths (brown, red), a less pronounced standard gap still occurs when the secondary becomes fully convective.

\clearpage

\centerline{\includegraphics[width=90mm]{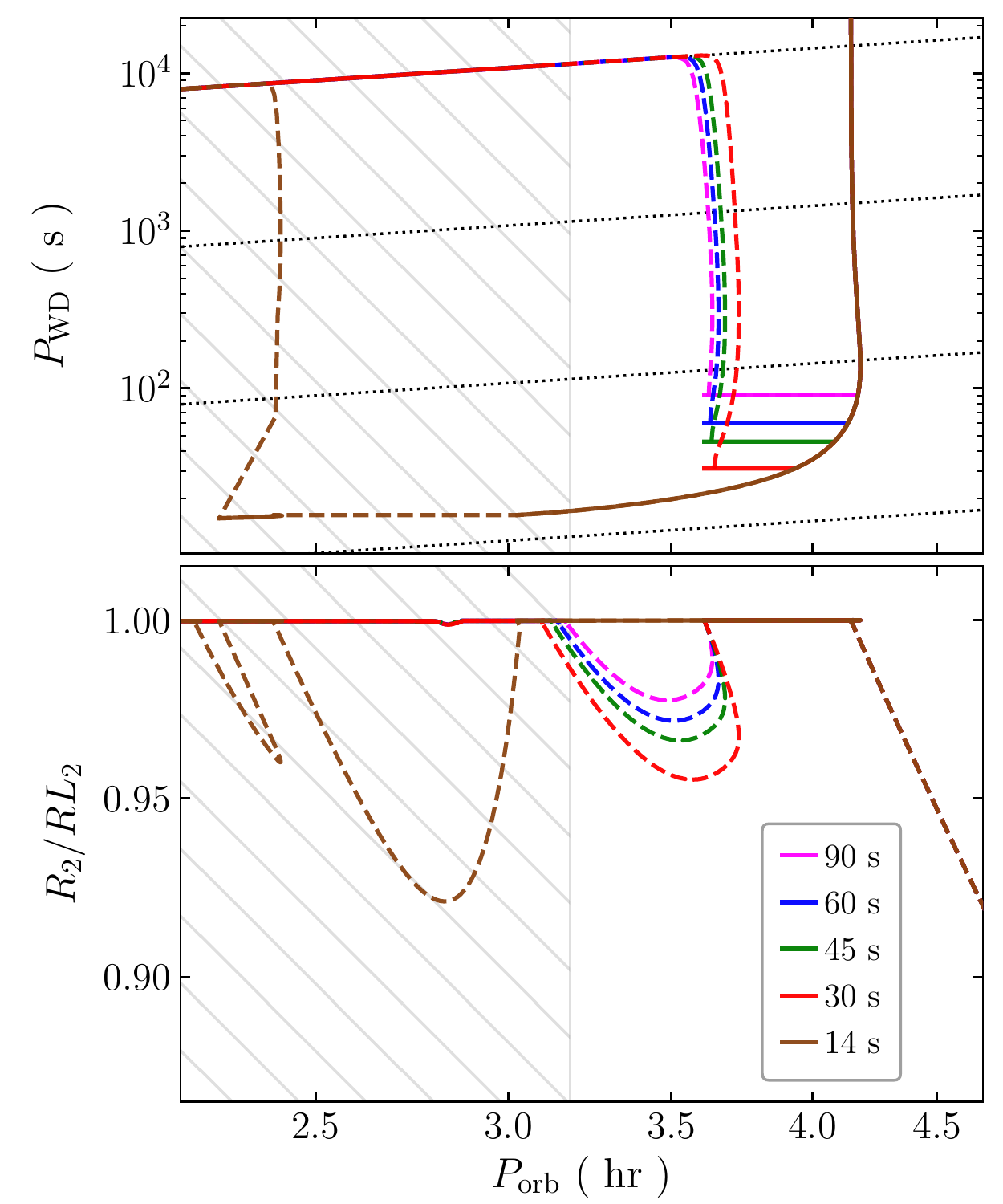}}
\noindent\textbf{Supplementary Figure\,4.} Evolutionary sequences highlighting the impact of 
different choices of the white dwarf spin 
period ($P_{\rm WD}$) at onset of magnetic field generation. The hatched region indicates the locus of the standard orbital period gap. 
In each track, solid and dashed lines correspond to semi-detached and detached phases respectively. 
With the exception of the case ${P_{\rm WD}=14}$\,s (brown curve), a detached phase is generated above the standard orbital period gap, at an orbital period $\approx3.75$\,hr. 
For $P_{\rm WD}$ at least twice the break-up period,
magnetic field generation starts at the same period and 
the slower the spin, the shorter the detached phase.
To spin-up the white dwarf to velocities close to the break-up period (brown track), the spin-up phase is much longer and therefore the pronounced detached phase occurs at a significantly shorter orbital period.



\end{bibunit}


\end{document}